\documentclass[12pt]{article}
\usepackage{amsmath}
\usepackage{graphicx}
\newcommand{\Tr}{{\rm Tr}\,}
\renewcommand{\Im}{{\rm Im}\,}
\newcommand{\be}{\begin{equation}}
\newcommand{\ee}{\end{equation}}
\newcommand{\bea}{\begin{eqnarray}}
\newcommand{\eea}{\end{eqnarray}}
\newcommand{\nl}{\nonumber \\}

\newcommand{\psibar}{\,\overline{\phantom{I}}\!\!\!\!\psi}

\newcommand{\Eq}[1]{{\rm Eq.\,}(\ref{#1})}
\newcommand{\eq}[1]{{\rm eq.\,}(\ref{#1})}

\newcommand{\refeq}[1]{(\ref{#1})}
\newcommand{\OO}{{\cal O}}
\newcommand{\gsim}{\stackrel{>}{{}_{\sim}}}
\newcommand{\lsim}{\stackrel{<}{{}_{\sim}}}
\newcommand{\mD}{m_D}
\renewcommand{\Im}{{\rm Im}\,}
\renewcommand{\Re}{{\rm Re}\,}

\begin{document}
\baselineskip=15.5pt
\pagestyle{plain}
\setcounter{page}{1}

\begin{titlepage}




\begin{center}

\vskip 1.7 cm

{\LARGE {O(g) Plasma Effects in Jet Quenching}}

\vskip 1.5cm
{\large 
Simon Caron-Huot$^{\dagger}$}

\vskip 1.2cm

$^{\dagger}$ Department of physics, McGill University,
Montr\'eal, Canada.

{\tt scaronhuot@physics.mcgill.ca}

\medskip
\vskip .5cm

\vskip 1.7cm

{\bf Abstract}
\end{center}

\noindent
We consider the bremsstrahlung energy loss
of high energy partons moving in the quark-gluon plasma, at weak coupling.
We show that the rates for these processes receive
large O(g) corrections from classical (nonabelian) plasma physics effects,
which are calculated.
In the high-energy (deep LPM) regime these corrections can be absorbed in
a change of the transverse momentum broadening coefficient $\hat{q}$,
which we give to the next-to-leading order.
The correction is large even at relatively
weak couplings $\alpha_s\sim 0.1$, as is typically found for such
effects, signaling difficulties with the perturbative expansion.
Our approach is based on an effective ``Euclideanization'' property of
classical physics near the light-cone,
which allows an effective theory approach based on dimensional
reduction and suggests new possibilities for
the nonperturbative lattice study of these effects.

\end{titlepage}

\newpage

\section{Introduction}

The phenomenon of jet quenching, or suppression of high-$p_T$ hadrons in $A+A$
collisions relative to expectations from
scaling of binary $p+p$ collisions,
has been the focus of much recent interest in RHIC physics
\cite{exp1} \cite{exp2}.
Its theoretical description (\cite{qinetalcomp} and references therein)
is based on the theory of jet evolution in thermalized media,
whose uncertainties it is thus worthwhile to seek to reduce,
or at least, quantify.
This requires the calculation
of higher-order effects,
which we propose to do in this paper in the regime of weak
coupling.

As established by a large body of work on the thermodynamic
pressure \cite{ArnoldZhai} \cite{braatennieto} \cite{gsixth},
finite temperature perturbation theory meets with serious convergence
difficulties.
Unless the strong coupling $\alpha_s$ obeys
$\alpha_s\lsim 0.1$,
strict perturbation theory in powers of $g$ is unreliable.
Such a behavior seems generic: it is also observed
for the next-to-leading order (NLO, $\OO(g)$)
corrections to thermal masses \cite{massnlo} \cite{massnlo1} \cite{blaizot00},
as well as for
the only transport coefficient presently known at NLO,
heavy quark momentum diffusion \cite{previouswork}
(whose behavior appears even worse).


Following Braaten and Nieto \cite{braatennieto}, who studied the
thermodynamic pressure,
these large perturbative corrections can be attributed
to purely classical (nonabelian) plasma effects. 
They have shown this by
first making use of the scale separation $gT\ll 2\pi T$ to integrate out
the scale $2\pi T$, leaving out a three-dimensional effective
theory (``electric QCD'', or EQCD)
describing the scale $m_D\sim gT$ as well as more infrared scales.
The claim is then that contributions from the scale $2\pi T$, as
well as the parameters of the effective theory,
enjoy well-behaved perturbative series \cite{braatennieto}
\cite{lainecoeffs}; all large corrections are included in
the effective theory.
Furthermore,  by treating this effective theory nonperturbatively
using various resummation schemes \cite{blaizot00} \cite{rebhan03} 
or the lattice \cite{kajantielattice},
reasonable convergence can be obtained down to $T \sim 3-5 T_c$.

It is natural to expect large corrections from $gT$-scale plasma effects
in other quantities as well.
Unfortunately, for real time quantities such as most transport
coefficients and collision rates, a resummation program
similar to that available in Euclidean space has
yet to be fully developed and applied.
This is because the real-time description of plasmas requires the hard thermal loop
(HTL) theory \cite{HTLrefs} (which in essence is classical (nonabelian) plasma
physics \cite{blaizot01}, also known as the
Wong-Yang-Mills system \cite{classical}),
which is arguably more complicated
than its Euclidean counterpart EQCD.



In this paper, we aim to point out
progress which can be made for a specific class of ``real-time''
quantities: those which
probe physics near the light cone.
This includes the collision kernel $C(q_\perp)$ that is relevant
for the transverse evolution of jets, whose crucial
role in the theory of jet quenching will be reviewed below.

To explain the idea, we first observe that the soft contribution
to $C(q_\perp)$ (that arising from soft collisions with
$q_\perp\sim gT$) is described by soft classical fields
that are being probed passively by the high-energy jet
passing through them.  These soft classical fields are the fields
surrounding the plasma particles.
At this point we observe that field components
moving collinearly with the jet are not particularly important ---
the standard calculation of collision rates 
\cite{braatenthoma} (or see \eq{dummyhard} below)
reveals that the contributing particles move with generic angles
in the plasma frame, with even a suppression for the ones collinear to
the jet (due to the reduced center-of-mass energy) ---
which implies that the result
must be insensitive to the precise value of the jet velocity $v\approx 1$.
The trick is then to think of $v$ as $v=1{+}\epsilon$ --- which, though
unphysical, doesn't affect the answer ---
thus making the hard particle's trajectory \emph{space-like}.
This makes Euclidean techniques directly applicable
including dimensional reduction, as will be explained below,
thereby dramatically simplifying the calculation.

In this paper we will thus
(analytically) compute the full $\OO(g)$ corrections to the
transverse collision kernel $C(q_\perp)$,
describing the evolution of the transverse momentum of a fast particle.
The second moment of that kernel gives the phenomenologically
interesting momentum broadening
coefficient $\hat{q}$, which we also compute at NLO.

This paper is organized as follows.
In section \ref{sec:res} we summarize our results
and explain their relevance to jet quenching; in particular
we discuss the relevance of the parameter $\hat{q}$.
In section \ref{sec:strategy} we explain our computational strategy
and formalism.
Details of the calculation of $C(q_\perp)$
and of its (ultraviolet-regulated) second moment $\hat{q}$
are given in sections \ref{sec:calc} and \ref{sec:qhat}, respectively.
In section \ref{sec:threepole} we derive, at NLO, the
relation between the collision kernel $C(q_\perp)$ for
momentum broadening and that for jet
evolution --- which turns out to be identical to the leading-order
relation --- and we discuss certain operator ordering issues
which could enter higher-order treatments.
Finally, in Appendix \ref{app:sumrule} we relate our approach
to a slight generalization of sum rules
previously found by Aurenche, Gelis and Zaraket \cite{AGZ}.

Alternative estimates of $\hat{q}$ and of jet evolution, based on
gauge-string duality
(see for instance \cite{rajagopal,rajarobust} \cite{teaney} \cite{gubser}
\cite{hattaiancu} \cite{gubsergluon} \cite{chesler}),
will not be discussed in this paper.

\section{Results}
\label{sec:res}

\subsection{Collision kernel}

The main result of this paper is the full next-to-leading order ($\OO(g)$)
(analytic) expression \refeq{finalres} for the two-body collision kernel
$C(q_\perp)$, defined as:
\be
\frac{d\Gamma}{d^2q_\perp/(2\pi)^2} \equiv C(q_\perp),
\label{defc}
\ee
describing the evolution of the
transverse momentum of a hard particle (with $E\gsim T$).

\begin{figure}
\begin{center}
\includegraphics[width=7.5cm]{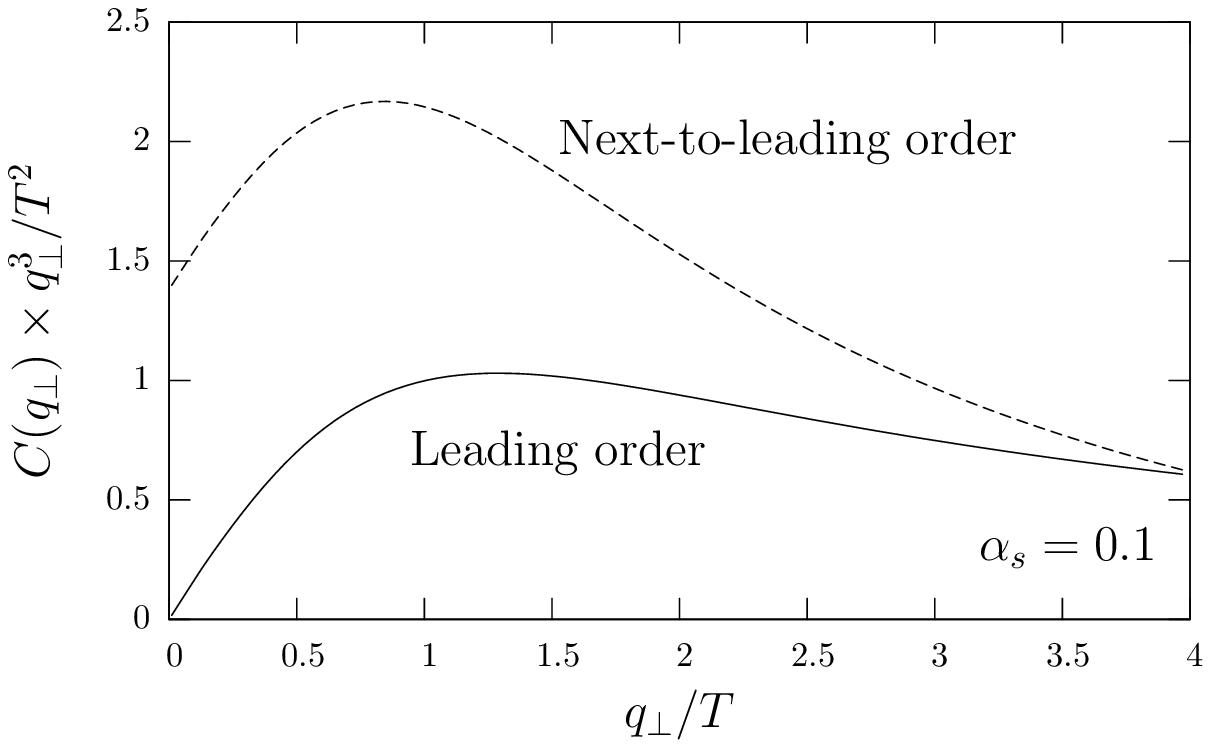}
\includegraphics[width=7.5cm]{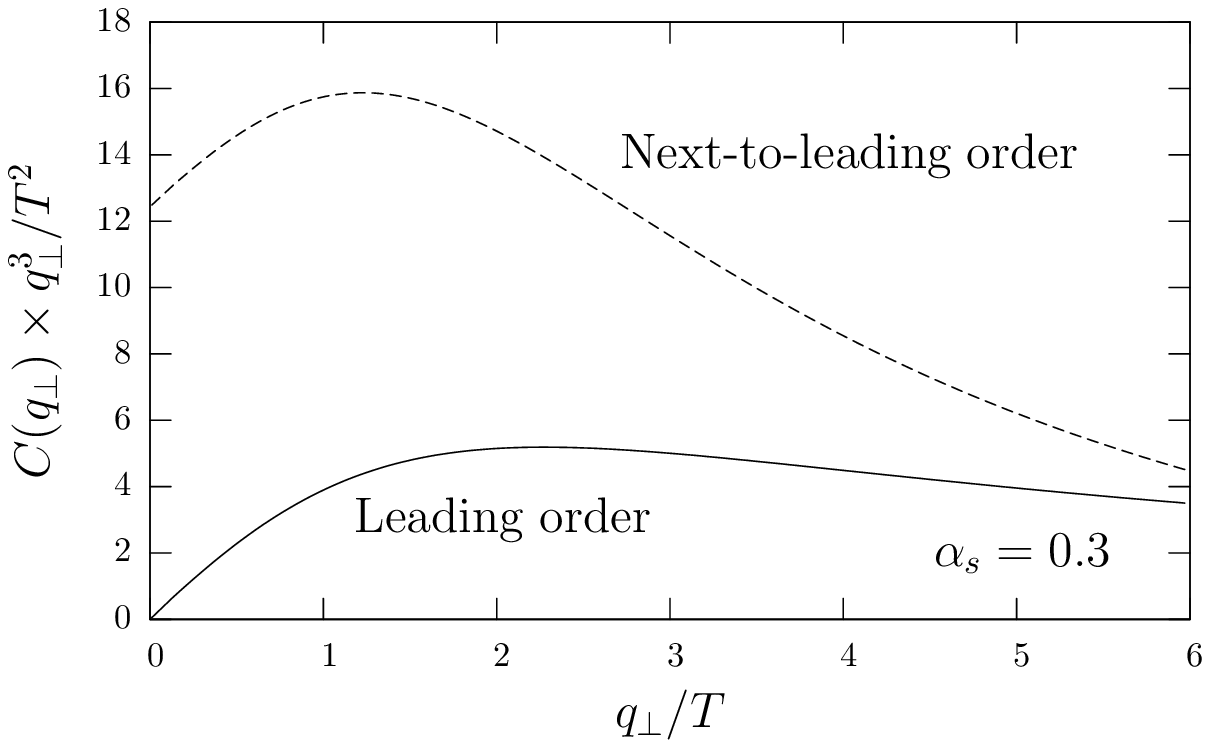}
\end{center}
\caption{LO and NLO collision kernels $C(q_\perp)\equiv(2\pi)^2 d\Gamma/d^2
  q_\perp$ for a fast quark in QCD (with $N_f=3$),
for $\alpha_s=0.1$ and $\alpha_s=0.3$.
For gluons the curves are to be multiplied by a (Casimir) factor $9/4$.}
\label{fig:coll}
\end{figure}

The $\OO(g)$ corrections to $C(q_\perp)$, which being due to $gT$-scale
physics only arise for $q_\perp\sim gT \ll T$,
are illustrated in fig.~\ref{fig:coll}.
Both the LO and NLO kernels $C(q_\perp)$ are proportional to the (quadratic)
Casimir of the gauge group representation of the jet.
The ``leading order curves'' is based on the full
(unscreened) expression \refeq{dummyhard} at hard momenta,
multiplied by $q_\perp^2/(q_\perp^2{+}m_D^2)$ to make it merge
smoothly with the analytic result \refeq{LOkernel} at low momenta,
following the prescription given in \cite{arnoldveryrecent}.
The ``next-to-leading order'' curves use the leading order curves
plus $C(q_\perp)^{(\rm NLO)}$ given in \refeq{finalres}.

The NLO correction is already quite large
for $\alpha_s=0.1$, giving nearly a factor of 2 around $q_\perp\approx T$.
As discussed in the Introduction, this
is consistent with the behavior observed for $\OO(g)$ effects in other
quantities.  At $\alpha_s=0.3$, a typical value used in comparisons
with RHIC data (see e.g. \cite{qinetal}),
it is clear that the strength of the correction has grown out of
control, meaning that (presently unknown)
yet higher-order corrections are most certainly also important
(though our results suggest that the value of $\alpha_s$ needed to fit
the data might be significantly smaller than the estimate of
\cite{qinetal}).

An interesting by-product of the approach
used in this paper is that it extends naturally to higher orders:
it makes perfect sense to evaluate the
gauge-invariant Wilson loop \refeq{dipole} nonperturbatively within the
\emph{Euclidean} three-dimensional EQCD theory,
for instance using the lattice.
Although this may not include \emph{all} $\OO(g^2)$ corrections to
$C(q_\perp)$ (contributions from the hard scale $2\pi
T$ will be missed),
by analogy with the works on the pressure discussed in the Introduction,
these missing contributions can be expected to be numerically suppressed%
 \footnote{Their description could turn out be very complicated,
   though, because jet evolution at $\OO(g^2)$ should
   contain, among other things, the analog of the NLO vacuum DGLAP splitting
   amplitudes in the presence of the LPM effect (described below).
   Also, various effects involving the scale evolution of the medium
   constituents and coupling constant evolution should arise.
 }.
We leave to future work the study of this interesting possibility.


\subsection{Application to Jet Evolution}
\label{sÂec:quench0}

The dominant energy loss mechanism of high energy particles
(at weak coupling) is bremsstrahlung (including
quark-antiquark pair production), triggered by soft collisions against
plasma constituents.
The theoretical description of these processes, at the leading order
in the coupling, is well-established \cite{BDMPS} \cite{zakharov}
\cite{AMYgluon}.
Their duration $t_{\rm form}$ depends on the energy of the
participants, and can interpolate between the Bethe-Heitler
(single scattering) regime
$t_{\rm form}\sim E/q_\perp^2 \sim E/m_D^2$ at energies $E\lsim T$,
and the Landau-Pomeranchuk-Migdal (LPM) \cite{LPM} (multiple-scattering)
regime at high energies $E\gg T$,
with $t_{\rm form}\sim \sqrt{E/\hat{q}}$, in which destructive
interference between different collisions plays a significant role.

In all of these regimes, however, the description factors
into a ``hard'' collinear splitting vertex
(Dokshitzer-Gribov-Lipatov-Altarelli-Parisi, DGLAP vertex \cite{DGLAP}),
times an amplitude (wavefunction in the transverse plane)
which describes the in-medium evolution of the vertex.
The latter accounts for the collisions which
trigger, and occur during, the splitting process
\cite{BDMPS} \cite{zakharov} \cite{AMYgluon}.
The DGLAP vertices themselves only involves hard scale physics
(in essence, they are Clebsch-Gordon coefficients between
states of different helicities) and thus cannot receive
$\OO(g)$ corrections; the NLO effects,
which come from soft classical
fields with $p\sim gT$, are included in their dressing amplitude.

In section \ref{sec:threepole} we discuss these amplitudes at NLO
and show that the relevant (three-body) collision kernel
factors as a sum of two-body kernels $C(q_\perp)$,
exactly like the LO one does
\cite{BDMPS} \cite{zakharov} \cite{AMYgluon,jeonmoore}.
As a consequence, our results can be used to give a full NLO
treatment of radiative jet energy loss;
one must simply include the NLO shift \refeq{finalres}
to the two-body kernel $C(q_\perp)$
which serves as an input to these calculations%
 \footnote{
   For instance, one would simply modify ``$C(q_\perp)$'' in
   \cite{jeonmoore},
   which is actually equal to $C(q_\perp)/(g^2C_s T)$ in our conventions.
 }.

\subsection{Momentum broadening coefficient ($\hat{q}$)}
\label{sec:broad}

When the effects of a large number of small collisions are added together,
it is natural to replace them by an effective diffusive process.
The diffusion coefficient relevant for transverse momentum
broadening, $\hat{q}$,
is defined as the second moment of the collision kernel \refeq{defc}:
\be
\hat{q}\equiv \int^{q_{\rm max}}_0
\frac{d^2q_\perp}{(2\pi)^2} q_\perp^2 C(q_\perp).
\label{defqhat}
\ee
The ultraviolet cutoff $|q_\perp|< q_{\rm max}$ is needed
to deal with the weak power-law falloff
$C(q_\perp)\sim g^4T^3/q_\perp^4$ at large $q_\perp$, which leads
to a logarithmic dependence of $\hat{q}$ on $q_{\rm max}$.
This is a \emph{leading order} logarithm;
below we shall comment on the value of the cutoff $q_{\rm max}$.
Using our NLO kernel \refeq{finalres} we can calculate the expansion
of $\hat{q}$ up to terms of order $g^2$ \footnote{Version 1 of the present paper
 contained a transcription error which is now corrected in \refeq{hardexpansion}.  The argument
 of the logarithm  (incorrectly) read $q^*$ as opposed to $\mD$
 obtained from \refeq{hardint} and \refeq{softint}.}:
\bea
\frac{\hat{q}}{g^4C_sT^3} \!\!&=& \!\!
\frac{C_A}{6\pi}\left[
\log \left(\frac{T}{\mD}\right)+
   \frac{\zeta(3)}{\zeta(2)} \log\left(\frac{q_{\rm max}}{T}\right)
-0.068854926766592\ldots\right]
\nl
&+&\!\! \frac{N_fT_f}{6\pi}\left[
\log \left(\frac{T}{\mD}\right) + \frac32
   \frac{\zeta(3)}{\zeta(2)} \log\left(\frac{q_{\rm max}}{T}\right)
-0.072856349715786\ldots\right]
\nl &+&
\frac{C_A}{6\pi} \frac{m_D}{T}\xi^{(\rm NLO)} +\OO(g^2)\,,
\label{hardexpansion}
\eea
with $\xi^{(\rm NLO)}=\frac{3}{16\pi}\left( 3\pi^2 + 10 -4\log
2\right)\simeq 2.1985$ a constant calculated in section \ref{sec:qhat},
characterizing the NLO correction to $\hat{q}$,
and $m_D^2=g^2T^2(N_c+N_fT_f)/3$ the leading-order Debye mass.
In QCD with $N_f=3$ flavors of fundamental quarks, $C_A=3$ and $N_fT_f=1.5$.
For a discussion of the leading order result and logarithms,
we refer the reader to P. Arnold's work \cite{arnoldveryrecent},
from which the high-precision numbers were taken.

The series \refeq{hardexpansion} is meant to represent the expansion in
$g$ of
the area under the curve of plots such as fig.~\ref{fig:coll}.
For $\alpha_s=0.1$ and a quark ($C_s=\frac43$) the area under the leading order curve in the figure
 (up to $q_{\rm max}=4T$) would yield
$\hat{q}^{\rm LO}\approx 2.60~ (\mbox{T/GeV})^3\, {\rm GeV^2/fm}$ 
 whereas the first two lines of \refeq{hardexpansion} give
$\hat{q}^{\rm LO, th} \approx 2.08 ~(\mbox{T/GeV})^3\, {\rm GeV^2/fm}$.
 The NLO shift is
$\Delta\hat{q}\approx 2.22 ~(\mbox{T/GeV})^3\, {\rm GeV^2/fm}$ from the figure, about a factor
 of two effect,
and $\Delta\hat{q}^{\rm th}\approx 5.26 ~(\mbox{T/GeV})^3\, {\rm GeV^2/fm}$ according to
\refeq{hardexpansion}.
Thus the third line of \refeq{hardexpansion} itself suffers from sizeable truncation errors
compared to our full NLO result \refeq{finalres}.
We would like to stress, however, that
\refeq{finalres}, and fig.~\ref{fig:coll},  is not merely
a simple truncation error from a lower-order
contribution but represents a genuine NLO effects.

As discussed in the preceding subsection, it would be premature to
attempt comparison of our $\hat{q}$ results with experimental data,
since it is clear that (yet unknown) higher-order corrections should
also be important at physically relevant couplings.
It is also worth noting
that different approximation schemes taking $\hat{q}$ as input
(this excludes the AMY scheme \cite{AMYgluon,jeonmoore}, which uses the
full $C(q_\perp)$)
when fitted to RHIC data,
tend to disagree rather significantly on its preferred value
\cite{qinetalcomp};  since a critical analysis of these approximations
lies beyond our scope, this simply means it is not completely clear
which experimentally-extracted value of $\hat{q}$ we should comparing
with.

It seems appropriate here to
recall some subtleties associated with the phenomenological
parameter $\hat{q}$, which do not arise if one instead works with the full
collision kernel $C(q_\perp)$.
First, the value of the cutoff $q_{\rm max}$ to be used in \refeq{hardexpansion} is
process-dependent:
since the $q_\perp{>}q_{\rm max}$ tail of $C(q_\perp)$ describes
collisions occurring on a finite rate\footnote{I am indebted to G.~D.~Moore for
 this discussion. } 
$\Gamma_{(q_\perp>q_{\rm max})}\sim
g^4T^3/q_{\rm max}^2$, weighting them
with $q_\perp^2$ in \refeq{hardint} ceases to make sense for
$\Gamma_{(q_\perp>q_{\rm max})}^{-1}\gsim t_{\rm jet}$, with $t_{\rm
  jet}$ the jet's lifetime, to be replaced with
a formation time $t_{\rm form}$ for bremsstrahlung pairs in the context
of energy loss calculations.
Therefore,
parametrically, one should set $q_{\rm max}\sim \sqrt{g^4T^3t_{\rm jet}}$.
For bremsstrahlung in the deep LPM regime,
$t_{\rm jet}\to t_{\rm form} \sim \sqrt{E/\hat{q}}$ so $q_{\rm max}\sim
g(ET^3)^{1/4}$ \cite{arnolddogan}%
   \footnote{
     This should be contrasted with the often-used
     kinematic cutoff $q_{\rm max}^2\approx ET$.
     }.

Second, the presence of the ultraviolet tail implies
that collisions having $q_\perp \sim q_{\rm max}$ (with $q_{\rm max}$ the
physical cutoff as determined above), which are intrinsically
non-diffusive, already contribute at the
next-to-leading logarithm order to bremsstrahlung rates.
That is, they contribute at $\OO(1)$
compared to the log-enhanced diffusive contribution
$\sim \log q_{\rm max}/m_D$ coming from $m_D\ll q_\perp\ll q_{\rm max}$.
Therefore, approximations based on diffusive physics lead, at best,
to expansions in inverse logarithms of the energy.
Such expansions
were systematically studied in \cite{arnolddogan},
with the conclusion that formulae to \emph{next-to-leading} logarithm
can be trusted at least when $E_{\rm jet}\gsim 10T$, with the inclusion
of the subleading term
(e.g. the constant under the logarithm) being mandatory.
None of the presently applied approaches which rely on $\hat{q}$
as a phenomenological parameter
presently includes such subleading logarithms (see
\cite{qinetalcomp,majumderchar} and references therein
for an overview of these approaches).


%
%
%
%


\section{Strategy: Space-like Correlators and EQCD}
\label{sec:strategy}

\def\v{\tilde{v}}
\def\gammat{\tilde{\gamma}}
In this section, we relate certain correlators at space-time
separation (more precisely, correlators supported on
space-like, and light-like, hyperplanes of the type $x^0=\v x^3$, $\v \leq 1$),
to Euclidean-signature correlators.
We will then apply to them the formalism of dimensional
reduction.

\subsection{Space-like correlators}

Field operators at space-like separated points (anti)commute
with each other and their correlator does not depend on
the operator ordering.
For two-point functions at vanishing time separation, a well-known
Euclidean representation holds \cite{kapusta}:
\bea
G^>_{ij}(t=0,{\bf x})&\equiv& \frac{1}{\Tr e^{-\beta H}} \Tr e^{-\beta H}
\OO_i({\bf x}) \OO_j(0)
\label{dummysum0}
=T\sum_n \int \frac{d^3{\bf p}}{(2\pi)^3}
e^{i{\bf p {\cdot}x}} G_E(\omega_n,{\bf p})\,,
\label{dummysum1}
\eea
with the sum running over the Matsubara frequencies $\omega_n=2\pi i
nT$, $\beta=1/T$,
and with $G^E_{ij}$ the Euclidean correlator of some
operators $\OO_{i,j}$ taken here to be bosonic.

In Lorentz-covariant theories, \refeq{dummysum1}
can be extended immediately to any correlator which is equal-time in a suitable
boosted frame.
Specifically, under a $z$-axis boost with velocity $\v$,
the thermal density matrix transforms to:
\be
e^{-\beta H} \to e^{-\gammat \beta (H'+\v P'{}^3)},
\label{torus1}
\ee
the primed quantities referring to quantities in the boosted frame;
$\gammat=\frac{1}{\sqrt{1{-}\v^2}}$.
The identification of $H'$ and $P'{}^3$ as the generators of time and space
translation
shows periodic identification
$x'{}^\mu=x'{}^\mu+i\gammat(\beta,-\v\beta,0_\perp)$ for
the geometry associated to \refeq{torus1},
with associated quantization condition on the ``Matsubara frequencies''
$p'{}^0+\v p'{}^3=2\pi i n T/\gammat$.
The spatial momentum $p'{}^3$ must be kept real:
it serves as a label for the physical states living on the
$x'{}^0=0$ hyperplane.
Thus only the frequency $p'{}^0$ is complex.
This determines the extension of \refeq{dummysum1} to equal-time
two-point functions in the boosted frame:
\be
G^>_{ij}(x'{}^0=0,{\bf x}') =
\frac{T}{\gammat}\sum_n
\int \frac{d^3{\bf p'}}{(2\pi)^3}
e^{i{\bf p' {\cdot}x'}} G_E(p'_n{}^0,{\bf p'}),
\hspace{1cm}
p'_n{}^0 = -\v p'{}^3 + 2\pi i n \frac{T}{\gammat}\,.
\ee

It will be convenient to boost this formula
back to the plasma rest frame, and to write it for general
space-time arguments ($\v=\frac{x^0}{x^3}$):
\bea
G^>_{ij}(x^0,{\bf x}')&=&
T\sum_n \int \frac{d^3{\bf p}}{(2\pi)^3}
e^{-i(p^0_nx^0-p^3_n x^3-p_\perp{\cdot}x_\perp)}
\,G^E_{ij}(p_n^0,p^3_n,p_\perp)\,,
\label{mainres}
\nl
p^0_n &=& 2\pi i nT,\hspace{1cm}
p^3_n =  p^3 + 2\pi i nT \frac{x^0}{x^3}\,.
\eea

\Eq{mainres} is the main result of this section.
It differs from \refeq{dummysum1} only due to
the imaginary part
of $p^3_n$, which, in fact, is required for the convergence of the
sum over $n$:  it prevents the Fourier exponential to contain
exponentially growing terms as opposed to pure phases.
\Eq{mainres} extends in a straightforward way to any
higher-point correlator supported on $(\frac{x^0}{x^3}=\v)$-type
hyperplanes:
one gets a summation-integration $\sum_n \int_p$, with the $p_n$ as in
\refeq{mainres}, for all external legs, subject to the usual restriction
of momentum conservation (and thus of ``$n$ conservation''),
as for equal-time higher-point correlators \cite{kapusta}.
The momenta running in loops must also be ``twisted'' like those in
\refeq{mainres}, e.g. $\Im p^3=\v \Im p^0$,
to reflect the boosted-frame origin of the formula%
 \footnote{
  Note that this gives an implicit dependence on the velocity $v$ to
   $G^E_{ij}(p_n^0,p^3_n,p_\perp)$ appearing in \refeq{mainres}.
 }.
This ensures that the imaginary
part of every momentum is time-like,
which is the natural domain of Euclidean physics.

We will be interested in the amplitudes of ultrarelativistic dipoles
moving with velocity $v=1$, e.g. $x^3=x^0$ (note $\v=1/v$ in general).
Our derivation of \refeq{mainres}
might seem compromised, since an ``infinite'' boost
with velocity $\v=1$ obviously doesn't exist.
However, a more careful look at the argument reveals that
the boost plays no important role: after all we
un-did it in the end.
All that is really important, is that we can imagine quantizing the system
along hyperplanes parallel to $\v$, and express the thermal density
matrix within these hyperplanes.
Since it is certainly possible to quantize a system along light
fronts, the result \refeq{mainres} must hold for $x^3=x^0$.
An alternative derivation of this, based on sum rules, is also given
in App.~\ref{app:sumrule}.

The attentive reader might complain that
setting $x^3=x^0$ in \refeq{mainres} corresponds to taking a
$v\searrow 1$ limit, whereas the physically relevant regime $v\nearrow 1$
lies beyond the reach of \refeq{mainres}.
Are we claiming that these limits are equivalent in general?
No%
 \footnote{
    In strongly coupled theories accessible to gauge-string duality,
    these two limits are known to be physically distinct.
    A calculation of $\hat{q}$ for a physical massive quark moving with $v<1$
    (in the sense of its momentum broadening coefficient)
    by Teaney and Casalderrey-Solana \cite{teaney},
    and by Gubser \cite{gubser},
    found a divergence $\hat{q}\sim (1-v^2)^{-1/4} \sqrt{\lambda}T^3$
    as $v\nearrow 1$.
    This calculation is valid for energies $E<M^3/\lambda T^2$,
    beyond which the coherence time of the force acting on the quark
    becomes of order the time scale of Langevin dynamics \cite{gubser};
    a time-independent description is then impossible.
    This suggests that $\hat{q}$ for $v<1$
    should depend on a cutoff time scale, as is the case at weak coupling.

    On the other hand, the $v\searrow 1$ limit has been studied
    by Rajagopal, Liu and Wiedemann \cite{rajagopal,rajarobust},
    by embedding Euclidean worldsheets into ${\rm AdS}_5$ space,
    and no divergences were met in this limit.
    It is thus qualitatively quite distinct.
}.
Our claim, explained in the introduction, is merely that for
\emph{classical} plasma physics effects they are equivalent --- to the extent
that the observable of interest is only passively probing a soft
classical background, only a phase-space suppressed fraction of which
propagates collinearly with the jet, this seems to be rather robust.
It is unclear whether this will remain true when quantum effects are
included (which will enter at $\sim g^2$), though,
because of collinear components present in the jet's own wavefunction.

\subsection{Dimensional reduction}
\label{sec:dimred}

Naturally, the contribution from soft physics (momenta $\sim gT$)
to sums like \refeq{mainres} is expected to be dominated by the $n=0$ mode.
We will thus begin by ``integrating out'' the modes
with $n\neq 0$.

First we claim that loop diagrams for which all external momenta
have $n=0$ are equal to the standard ones.
These two sets of diagrams have $p^0=0$ and $p_z$ real, and
could only differ due to the ``twisted'' Matsubara momenta
\refeq{mainres} which circulate in the former; our claim is that
this does not affect their value.
The reason is that the imaginary part of every momentum $P$ in such
loops is time-like with a real part obeying $\Re p^0=0$,
ensuring that $\Re P^2$ is positive-definite.
The imaginary part of the $p^3$ integration contours can thus
be deformed from $\Im p^3= \v\Im p^0$ to $\Im p^3=0$,
without crossing any pole.

In particular, the modes with $n=0$ are described by precisely the
standard ``electric QCD'' (EQCD) three-dimensional effective theory
\cite{dimred0,dimred}.  EQCD is
pure three-dimensional Yang-Mills with coupling constant $g^2_3=g^2T$
coupled to a massive adjoint scalar $A_0$ of mass $m_D$.
It is an effective theory for the $gT$ scale (the Euclidean version of
the hard thermal loop theory \cite{HTLrefs}),
in which the loop expansion proceeds in powers of $g^2T/m_D\sim g$.
Its parameters do not receive $\OO(g)$ corrections.

The propagators of EQCD are:
\def\G{\tilde{G}}
\be
\G^{00}(q)=\frac{-1}{q^2+m_D^2},\hspace{1cm}
\G^{ij}(q)=\frac{\delta^{ij}}{q^2} - \frac{\xi q^iq^j}{q^4}. \label{propagators}
\ee
(We use the tilde to denote that these are three-dimensional
propagators.)
The minus sign in front of the $A^0$ propagator
reflects the fact that we will couple it to Minkowski-space
Wilson lines: we have \emph{not} performed a Wick rotation.

In addition to its interaction with the $n=0$ modes, we must also
include the direct coupling of the operator of interest to the
$n\neq 0$ modes.
Physically, and as shown in Appendix \ref{app:sumrule},
a contribution from these modes would correspond, in the real-time
formalism, to a failure of the soft approximation $n_B(p^0)\approx T/p^0$.
Such a failure would signal a contribution from the $p^0\sim T$ region
in Minkowski space, which would necessary be signaled by
ultraviolet divergences in the soft approximation,
since this approximation correctly describes
the intermediate region $gT\ll p^0\ll T$ and any contribution
from the scale $T$ should leave an imprint on this region.
Thus, provided we do not find ultraviolet divergences from the $n=0$
contribution alone (which computes exactly the soft approximation,
see Appendix ~\ref{app:sumrule}),
this argument shows that we
can safely ignore the direct coupling to the $n\neq 0$ modes.
This will turn out to be our case.

\section{The calculation}
\label{sec:calc}

In this section we express the collision kernel $C(q_\perp)$ as a
correlator supported on $x^3=x^0$ trajectories and evaluate
it using EQCD.
We only give details in the Feynman gauge $\xi=0$,
though we have explicitly verified the $\xi$-independence
of our (gauge-invariant) collision kernel, as a check on the
calculation.

\subsection{Operator definition of $C(q_\perp)$ and leading order result}

\begin{figure}
\begin{center}
\includegraphics[width=8cm]{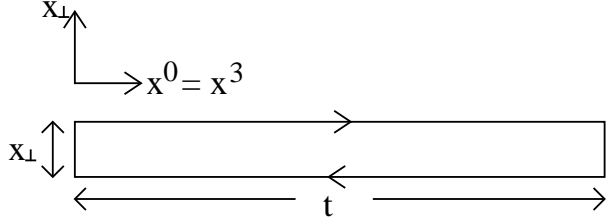}
\end{center}
\caption{Wilson loop representation of the dipole amplitude.}
\label{fig:loop}
\end{figure}

The evolution of the transverse momentum of a high-energy particle can be described
by looking at its density matrix, as described in detail in
\cite{teaney}.
For classical effects, however
(and even more so because we are taking the velocity to be $v=1{+}\epsilon$),
we can neglect operator ordering issues and replace the evolution
of the density matrix by that of a dipole;
we will come back to operator ordering issues in section
\ref{sec:oporder}.

High energy dipoles ($E\gg \mD$) propagate eikonally
in the soft classical background.
The collision kernel describing
the evolution of its transverse momentum can thus
be recovered from the Fourier transform of the (long-time limit of the)
dipole propagation amplitude $W$ \cite{BDMPS} \cite{zakharov} \cite{xnwang}:
\bea
W(t,x_\perp)&\sim& e^{-t C(x_\perp)+ \OO(1)},\, t\to\infty,
\label{dipole}
\nl
\Leftrightarrow
C(q_\perp)&\equiv& \int d^2x_\perp e^{ip_\perp{\cdot}x_\perp} C(x_\perp).
\eea
$C(q_\perp)$ is short for $(2\pi)^2d\Gamma/d^2q_\perp$, as in
\refeq{defc}.
The dipole amplitude
$W(t,x_\perp)$ is given by the trace of
a long, thin rectangular
Wilson loop stretching along the light-cone coordinate
$x^+$, with a small transverse extension $x_\perp$ (see fig.~\ref{fig:loop}).

The naive dimensional reduction of the Wilson loop
\refeq{dipole} yields a Wilson loop
stretching along the $z$-axis of the three-dimensional EQCD theory.
It couples to the linear combination $A_+\equiv (A_z+A_0)$ of the EQCD
fields, reflecting its ultrarelativistic origin.
This ``naive'' dimensional reduction corresponds to keeping only
the direct coupling to the $n=0$ modes.  As explained in
subsection \ref{sec:dimred}, this will be justified provided
we do not find ultraviolet divergences.

\begin{figure}
\begin{center}
\includegraphics[width=15cm]{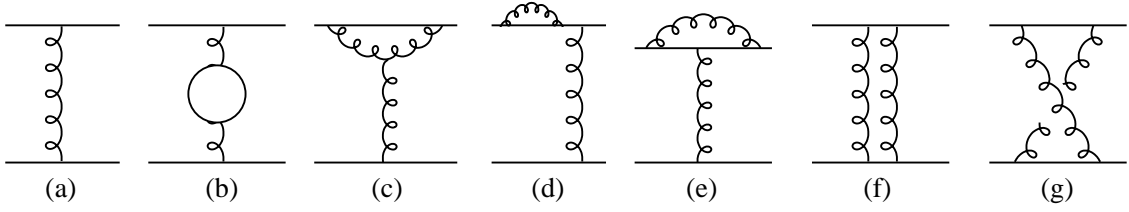}
\end{center}
\caption{Tree and one-loop diagrams contributing to
$C(q_\perp)$.}
\label{fig:diags}
\end{figure}

At the lowest order in perturbation theory, only
the single-gluon exchange diagram ((a) of fig.~\ref{fig:diags}) contributes,
\bea
C(q_\perp)&=& g^2TC_s\int_{-\infty}^\infty dz \int d^2x_\perp
e^{ip_\perp{\cdot}x_\perp} \G_{++}(z,x_\perp)
\nl
&=&
g^2TC_s \G_{++}(q_z=0,q_\perp)
= g^2TC_s \left(\frac{1}{q_\perp^2}-\frac{1}{q_\perp^2{+}m_D^2}\right),
\label{LOkernel}
\eea
where we have used \refeq{propagators}.
The compact form \refeq{LOkernel} was first obtained by means of sum rules by
Aurenche, Gelis and Zaraket \cite{AGZ},
which we show in Appendix \ref{app:sumrule} are equivalent
to our approach.

\subsection{Diagram (b)}

\def\qp{(q_\perp{-}p)}

At the next-to-leading order (one-loop),
self-energy insertions to the single-gluon diagram (b) contribute
(we will often write ``$q_\perp$'' for a three-vector with $q_z=0$,
which should cause no confusion; ``$\int_p$'' is short for $\int \frac{d^3p}{(2\pi)^3}$):
\bea
C(q_\perp)_{(b)}/g^2TC_s &=& \frac{\delta\Pi^{00}(q_\perp)}{(q_\perp^2{+}m_D^2)^2}
- \frac{\delta\Pi^{zz}(q_\perp)}{q_\perp^4},
\nl
\frac{\delta\Pi^{00}(q)}{g^2TC_A}&=& -\int_p
\left[\frac{(2q_\perp-p)^2}{p^2(\qp^2+m_D^2)}
- \frac{3}{p^2} \right],
\nl
\frac{\delta\Pi^{zz}(q)}{g^2TC_A}&=& -\int_p
\left[\frac{ 2p_z^2 }
{(p^2{+}m_D^2)(\qp^2{+}m_D^2)}  - \frac{1}{p^2{+}m_D^2}\right]
\nl &&
-\int_p\left[ \frac{3p_z^2 +2 q_\perp^2 + p^2}{p^2\qp^2} - \frac{2}{p^2}
-\frac{p_z^2}{p^2\qp^2}\right].
\label{dummyse}
\eea
Each bracket includes the contributions of one fish and
one tadpole diagram, while the last one also includes the ghost
loop.

The (linear) ultraviolet divergences in
\refeq{dummyse} are to be canceled by matching counter-terms
that can be unambiguously calculated within the framework
of dimensional reduction \cite{dimred0,dimred}.
They merely represent the (hard thermal loop)
coupling of the $n\neq 0$ gluons to the soft $n=0$ ones,
e.g. the gluon contribution to the $A^0$ mass squared $m_D^2$.
The fact that the direct coupling to exchange gluons with
$q^0=q^3\neq 0$
does not contribute to the divergences can also be checked explicitly,
from the convergence, with respect to $q^3$, of the real-time integral
\refeq{dummyhard} (this justifies making the soft approximation on $q^0$).
Thus the divergences in \refeq{dummyse} do not signal the presence of ``new
contributions'' beyond the EQCD effective theory, as discussed in
section \ref{sec:dimred}.

Employing dimensional regularization, the divergences
simply go away%
\footnote
  {
    \label{foot:pole}
    The dimensionally-regulated
    integrals \refeq{dummyse} have poles in dimensions 2 and 4 but are
    finite and unambiguous in dimension 3.
  }
and the counter-terms are zero to $\OO(g)$ \cite{dimred}.
This way we obtain (all our arctangents run from $0$ to $\pi/2$):
\be
\frac{C(q_\perp)_{(b)}}{g^4T^2 C_s C_A} =
\frac{  {-}m_D-2\frac{q_\perp^2{-}m_D^2}{q_\perp}
  \tan^{-1} \left(\frac{q_\perp}{m_D}\right) }
{4\pi(q_\perp^2{+}m_D^2)^2}
+ \frac{7}{32 q_\perp^3}
+\frac{m_D-\frac{q_\perp^2{+}4m_D^2}{2q_\perp}
  \tan^{-1}\left(\frac{q_\perp}{2m_D}\right)}
{8\pi q_\perp^4}
\ee

\subsection{Diagram (c)}

Diagram (c) plus its permutation contribute:
\bea
\frac{C(q_\perp)_{(c)}}{g^4T^2C_s C_A} &=& \int_p
\left[
\frac{2}{q_\perp^2 (p^2{+}m_D^2) (\qp^2{+}m_D^2)}
-\frac{1}{(q_\perp^2{+}m_D^2) p^2 (\qp^2{+}m_D^2)}
\right.
\nl && \hspace{2cm} \left.
\label{dummyc}
-\frac{1}{(q_\perp^2{+}m_D^2) (p^2{+}m_D^2)\qp^2} \right]
\\
&=&
\frac{{-}\tan^{-1} \left(\frac{q_\perp}{m_D}\right) }
  {2\pi q_\perp(q_\perp^2+m_D^2)}
+\frac{\tan^{-1} \left(\frac{q_\perp}{2m_D}\right) }{2\pi q_\perp^3}
%
%
\label{resc}
\eea
In the Feynman gauge, there is no contribution involving only transverse
gauge fields: such a contribution would involve the
(trivial) $zzz$ vertex.
\Eq{dummyc} is manifestly convergent.

\subsection{Diagrams (d)-(g)}
\label{sec:diagsdg}

Our calculation is based on a quasiparticle expansion, e.g. we
simply set on-shell the external legs of scattering diagrams.
The relevant expansion parameter is $g$,
e.g. the ratio of the scattering width $\sim g^2T$
to the scattering's natural frequency scale $m_D$.
Thus in evaluating the external state corrections
(d) we need only keep those effects which are not suppressed by the
smallness of the width.
A narrow resonance being described by just its position and the total
area under it, this means that diagram (d), at $\OO(g)$, produces only
mass-shell corrections and wave-function renormalization factors.
The (here imaginary) ``mass-shell'' corrections have no effects:
they are identical for the initial and final states, so
the ``energy'' (read $z$-momentum) transfer is zero in any case.
The wave-function renormalization contribution is given by
an energy derivative of the eikonal self-energy,
and (e) is unambiguous, yielding
respectively (including all diagrams of similar topology):
\bea
C(q_\perp)_{(d)} &=& 2g^4T^2 C_s^2 \G_{++}(q_\perp) \int_p \G_{++}(p)
\frac{d}{dp_z} \frac{1}{p_z-i\epsilon}
\nl
C(q_\perp)_{(e)} &=& 2g^4T^2 C_s(C_s-\frac12 C_A)
\G_{++}(q_\perp) \int_p \G_{++}(p)
\frac{1}{(p_z-i\epsilon)^2}
\eea
The sum of (d) and (e) is proportional to $C_A$ and
identically vanishes in the abelian theory ($C_A=0$),
as required by abelian exponentiation%
\footnote
   {
   Abelian Wilson loops, computed using Gaussian distribution
   for gauge fields (as is done by diagrams (d)-(g), for which
   only the two-point function of the gauge field enters), simply
   exponentiate:
   $\langle e^{\int A}\rangle = \exp( \frac12 \langle\int A \int A \rangle)$.
   As a consequence, the collision kernel as defined from \refeq{dipole}
   is tree-level exact in such theories;
   there is no interference between scattering events.
  }.

Part of diagram (f) is already included
by the exponentiation of \refeq{LOkernel} (diagram (a)):
this generates the approximation
to (f) in which the intermediate eikonal propagators
are put on-shell.  To avoid double-counting, this must be subtracted.
We must first regulate the associated ``pinching'' ($q_z\to 0$) singularity,
which we do by flowing a small external $z$-momentum $\omega$
into the Wilson loop.  We then take the limit $\omega\to 0$ after the subtraction.
Diagram (g) poses no difficulty.
\bea
C(q_\perp)_{(f)}&=&
g^4T^2 C_s^2 \int_p \G_{++}(p) \G_{++}(q{-}p)
\left[\frac{1}{(p_z+i\epsilon)(p_z+\omega-i\epsilon)}
+ \frac{2\pi i\delta(p_z)}{\omega-i\epsilon}\right]_{\omega\to 0}
\label{diagf}
\\
C(q_\perp)_{(g)}&=& -g^4T^2 C_s(C_s-\frac12C_A)\int_q \frac{\G_{++}(p)
  \G_{++}\qp}{(p_z-i\epsilon)^2}.
\label{diagg}
\eea
\Eq{diagf} has a well-defined $\omega\to 0$ limit, as follows
from the identity $1/(p_z{+}i\epsilon-1/(p_z{-}i\epsilon)=-2\pi i
\delta(p_z)$.  This limit takes a form identical to \refeq{diagg}
and the sum is proportional to $C_A$,
again as required by abelian exponentiation.
This confirms the correctness of our evaluation of (f).

The sum of diagrams (d)-(g) yields:
\bea
\frac{C(q_\perp)_{(d)\ldots(g)}}{g^4T^2C_sC_A}&=&
\frac12 \int_p
\frac{
  \G_{++}(p)\G_{++}\qp - 2\G_{++}(p)\G_{++}(q_\perp)}{(p_z-i\epsilon)^2}
\label{dummydg}
\\
&=& \frac{m_D}{4\pi(q_\perp^2{+}m_D^2)}\left[
  \frac{3}{q_\perp^2{+}4m_D^2}
 -\frac{2}{(q_\perp^2{+}m_D^2)}
 -\frac{1}{q_\perp^2} \right]
\eea
The function $\G_{++}$ is $\G_{00}+\G_{zz}$,
as given in \eq{propagators}.
To evaluate \refeq{dummydg} we have found convenient to first apply integration
by parts to the $1/(p_z-i\epsilon)^2$ denominator, which removes the
  explicit $p_z$-dependence and reduces the integral to a set of standard
  isotropic Feynman integrals.
\Eq{dummydg} is manifestly infrared- (and ultraviolet-) safe,
upon enforcing $p\leftrightarrow \qp$ symmetry.




\subsection{Final formulae}

In summary, we have obtained all $\OO(g)$ contributions to the collision
kernel $C(q_\perp)$:
\bea
C(q_\perp)^{(\rm LO)} &=& \frac{g^2TC_s m_D^2}{q_\perp^2(q_\perp^2{+}m_D^2)}
\nl
\frac{C(q_\perp)^{(\rm NLO)}}{g^4T^2C_sC_A} &=&
\frac{7}{32 q_\perp^3} +
\frac{  {-}m_D-2\frac{q_\perp^2{-}m_D^2}{q_\perp}
  \tan^{-1} \left(\frac{q_\perp}{m_D}\right) }
{4\pi(q_\perp^2{+}m_D^2)^2}
+\frac{m_D-\frac{q_\perp^2{+}4m_D^2}{2q_\perp}
  \tan^{-1}\left(\frac{q_\perp}{2m_D}\right)}
{8\pi q_\perp^4}
\nl
&&
- \frac{\tan^{-1} \left(\frac{q_\perp}{m_D}\right) }
  {2\pi q_\perp(q_\perp^2+m_D^2)}
+\frac{\tan^{-1} \left(\frac{q_\perp}{2m_D}\right) }{2\pi q_\perp^3}
\nl
&& +
\frac{m_D}{4\pi(q_\perp^2{+}m_D^2)}\left[
  \frac{3}{q_\perp^2{+}4m_D^2}
 -\frac{2}{(q_\perp^2{+}m_D^2)}
 -\frac{1}{q_\perp^2} \right].
\label{finalres}
\eea
These expressions are valid for $q_\perp \ll T$.
The leading order kernel for $q_\perp \gsim T$ gets
slightly modified, see \refeq{dummyhard}.

The reader might wonder as to the appearance of arctangents
with two distinct arguments in \refeq{finalres}.
They can be understood by looking in the complex $q_\perp^2$-plane:
$\tan^{-1}(q_\perp/2m_D)$
has a branch cut starting at $q_\perp^2=-4m_D^2$,
exhibiting its origin from the
exchange of a pair of two quanta of mass $m_D$
(longitudinal gluons).
The branch cut of $\tan^{-1}(q_\perp/m_D)$, starting at
$q_\perp^2=-m_D^2$, arises from the exchange of
one longitudinal and one transverse gluon.
Both arctangents occur, since both of these pairs of states
can be exchanged.  Exchange of two massless quanta also occurs,
and generates $1/\sqrt{q_\perp^2}$-type of discontinuities instead of
arctangents.

\section{Evaluation of $\hat{q}^{(\rm NLO)}$}
\label{sec:qhat}

The effective theory approach we have used so far is valid
for $q_\perp \ll T$.  As mentioned in section \ref{sec:broad},
however, the momentum broadening coefficient
$\hat{q}$ (second moment of $C(q_\perp)$)
receives contributions from all scales up to a
process-dependent cut-off $q_{\rm max}$.
In this section we will assume%
\footnote{
 In the context of jet quenching this is (parametrically) equivalent to
 $E\gg T/g^4$, with $E$ the \emph{smallest} energy of the participants.
} $q_{\rm max}\gg T$.

To separate the soft and hard contributions to $\hat{q}$,
we find convenient to introduce an auxiliary scale $q^*$
obeying $m_D\ll q^* \ll T$:
\be
\hat{q} =
\int_0^{q^*} \frac{d^2q_\perp}{(2\pi)^2} q_\perp^2 C(q_\perp)^{\rm soft}
+
\int_{q^*}^{q_{\rm max}}
\frac{d^2 q_\perp}{(2\pi)^2} q_\perp^2 C(q_\perp)^{\rm hard}
\label{qhatint}
\ee

The soft kernel $C(q_\perp)^{\rm soft}$ is given by \refeq{finalres}.
The hard kernel $C(q_\perp)^{\rm hard}$ describes tree-level $2\to 2$
scattering processes against plasma constituents,
with self-energy corrections omitted on the exchange gluon
(since they represent only $\sim g^2$ corrections for $q_\perp \sim T$).
The large particle energy $E\gg T$ guarantees that the Mandelstam invariants
$s\sim ET$ and $-t=q_\perp^2$ obey $|t|\ll s$, so that the
relevant scattering matrix elements
assume the universal (eikonal) form $\propto s^2/t^2$.
The kinematics force $q^0=q_z$ for the momentum transfer $q$.
In fact, these processes are precisely described by the central cut of the
(four-dimensional) diagram (b) of fig.~\ref{fig:diags}.
Performing the $q_z$ integration in the expression for the
collision rate (as done in \cite{braatenthoma}; more details
can be found in \cite{arnoldveryrecent}), one obtains:
\be
C(q_\perp)^{\rm hard} = \frac{g^4C_s}{q_\perp^4} \int \frac{d^3p}{(2\pi)^3}
\frac{p-p_z}{p}
\left[ 2C_A n_B(p)(1+n_B(p')) + 4N_f T_f n_F(p)(1-n_F(p'))\right],
\label{dummyhard}
\ee
with $p,p'$ the initial and final momentum of the target particle;
$p'=p+q_z$, $q_z=q^0=\frac{q_\perp^2+2q_\perp{\cdot}p}{2(p-p_z)}$.
In the regime $q_\perp\ll T$, $p'\approx p$ and
\refeq{dummyhard} reduces (as it must) to the
large $q_\perp$ limit of \refeq{LOkernel},
$C(q_\perp)\approx g^2m_D^2C_sT/q_\perp^4$.

Integrating \refeq{dummyhard} over $q$ to obtain the hard contribution
to \refeq{qhatint}, and expanding it in powers of $q^*/T$, yields:
\bea
\frac{\hat{q}^{\rm hard}}{g^4C_sT^3} \!\!&=& \!\!
\frac{C_A}{6\pi}\left[
\log \left(\frac{T}{q^*}\right)+
   \frac{\zeta(3)}{\zeta(2)} \log\left(\frac{q_{\rm max}}{T}\right)
-0.068854926766592\ldots + \frac{3}{16} \frac{q^*}{T} +\ldots \right]
\nl
&+&\!\! \frac{N_fT_f}{6\pi}\left[
\log \left(\frac{T}{q^*}\right) + \frac32
   \frac{\zeta(3)}{\zeta(2)} \log\left(\frac{q_{\rm max}}{T}\right)
-0.072856349715786\ldots + \ldots\right]\!\!,
\label{hardint}
\eea
with the omitted terms being suppressed by $(q^*/T)^2$ or more.
The quoted numbers are from \cite{arnoldveryrecent}; we have verified
the first five significant digits by direct numerical integration
of \refeq{dummyhard}%
 \footnote{I thank P.~Arnold for pointing to me a numerical error in an early
   draft of this paper.
 }.
The $\sim q^*/T$ term arises from soft bosons
with $p,p'\ll T$ and can be obtained in
the soft approximation $n_B(p),n_B(p')\to T/p,T/p'$; it is also given
in \cite{arnoldveryrecent}.

The soft contribution to \refeq{qhatint}, e.g. the second moment of
\refeq{finalres}, admits the expansion:
\bea
\hspace{-0.7cm} \frac{\hat{q}^{\rm soft}}{C_s} &=&
\frac{m_D^2g^2T}{2\pi} \log\left(\frac{q^*}{m_D}\right) +
\frac{g^4T^2 C_A m_D}{2\pi} \left[
{-}\frac{q^*}{16 m_D}
+ \frac{3\pi^2 + 10 -4\log 2}{16\pi}  {+} \ldots
\right]\!\!,
\label{softint}
\eea
with the omitted terms being suppressed by powers of $m_D/q^*$.
The $q^*$ dependence of \refeq{hardint} and \refeq{softint} cancels
out in their sum, as it must do, producing
the advertised formula \refeq{hardexpansion}.
This cancellation provides a rather nontrivial check on the calculation.

The reader might inquire as to whether we have consistently included
all $\OO(g)$ contributions to $\hat{q}$.
Taking $q^*\sim g^{1/2}T$, for instance, the omitted terms
$\sim (q^*/T)^2$ in \refeq{hardint} might naively appear to be $\OO(g)$,
suggesting contributions from other, omitted terms.

Estimates of this kind can be misleading, however, because
$q^*$ is not a physical scale in this problem.
The matching region $m_D\ll q^*\ll T$ can be described
equivalently using the low-energy description (EQCD) or
the full theory, ensuring that
$q^*$ always disappears from final expressions.
This is seen explicitly for the leading truncation errors $\sim q^*/T$
in \refeq{hardint} and \refeq{softint}: instead of producing
$\OO(g^{1/2})$ corrections, as one would naively expect setting
$q^*\sim g^{1/2}T$,
they cancel against each other
and the leading correction is $\OO(g)$, not $\OO(g^{1/2})$.
Since similar cancellations are bound to occur at all orders,
this simply
means that the scale $q^*$ should not enter power-counting estimates.
Because higher loop diagrams are $\sim g^2$ when $q_\perp\sim T$ and
because we have included all $\OO(g)$ effects when $q_\perp \sim m_D$,
we thus conclude that we have included all $\OO(g)$ contributions.

Finally, we note that, in the spirit of \cite{DRnice},
we could have used
dimensional regularization to separate the $q$ integration,
instead of the sharp cutoff $q^*$.
In this scheme, the hard $q^*/T$ term in \refeq{hardint} disappears:
there is no suitable dimensionful parameter to replace $q^*$.
The $\OO(g)$ corrections then come solely from the
(unambiguous) dimensionally-regulated soft integral \refeq{finalres}.

\section{Jet Evolution}
\label{sec:threepole}

We now extend the calculation to obtain the collision kernel
relevant
for bremsstrahlung and pair production processes.
The new complication is that, except for QED processes,
the relevant object to evolve in the plasma is no longer a ``dipole'':
it involves three charges.
For instance, to describe the gluon bremsstrahlung process $\psi\to
g\psi$, one must evolve an operator which annihilates a quark
and creates a quark-gluon pair (see \cite{BDMPS} \cite{zakharov}
\cite{AMYgluon}, which, however, use somewhat different notations):
\be
\OO_{\psi\to\psi g} = |\psi,g\rangle\langle \psi|\,.   \label{threepole}
\ee
The three color charges in \refeq{threepole}
are paired together to form a color-singlet state, as
dictated by the (DGLAP) gluon emission vertex
which generates this operator.

It turns out that only one transverse momentum suffices
to describe the internal state of \refeq{threepole}.
A priori the description of three charges might seem to require
three momenta,
but momentum conservation reduces this by one,
and a symmetry removes yet another one:
by suitably choosing the $z$-axis
it is always possible to ``gauge'' to zero one of the transverse momenta
(see the discussion preceding eq.(6.6) in
\cite{AMYgluon}%
  \footnote{
    For high-energy jets (when at least \emph{one} of the energy of the
    participant is large, $E_{\rm max}\gg T$), these rotations can be taken
    to have energy-suppressed angles $\sim q_\perp/E$, and thus to have
    negligible effects on the longitudinal momenta.
    Even when $E_{\rm max}\sim T$, the angles are at most $\sim g$ and the
    changes in longitudinal momenta are $\sim g^2$, beyond the accuracy
    considered in this paper.}).
In the following, for concreteness,
we shall gauge to zero the transverse momentum of particle 1,
and $q_\perp$ will refer to the transverse momentum of particle 2.

At the leading order, the relevant collision kernel is a sum over two-body
contributions \cite{BDMPS} \cite{zakharov} \cite{AMYgluon}:
\def\C{\tilde{C}}
\be
\frac{d\Gamma_3(q_\perp)}{d^2 q_\perp/(2\pi)^2} =
 \frac{C_2{+}C_3{-}C_1}{2} \C(q_\perp)
+\frac{C_1{+}C_3{-}C_2}{2} \C(\frac{E_1}{E_2}q_\perp)
+\frac{C_1{+}C_2{-}C_3}{2} \C(\frac{E_1}{E_3}q_\perp)
\label{gamma3}
\ee
with $C_i$ and $E_i$ respectively the Casimir and longitudinal momenta
of the participating particles; $\C(q_\perp)\equiv
\frac{1}{C_s}C(q_\perp)$ denotes a single-particle collision kernel
with its Casimir factor stripped off; we recall that the LO (and NLO)
kernels respect Casimir scaling.
In the special limit in which one of the $E_i$ becomes much
smaller than the other ones, the motion of this particle
dominates and the kernel \refeq{gamma3} reduces to the one for
single-particle diffusion,
$C(q_\perp)$, for $i=2,3$,
and $C(\frac{E_1}{E_2}q_\perp)$ when $i=1$.

As we presently show, it turns out that the formula
\refeq{gamma3} also holds
at NLO, provided the NLO expression \refeq{finalres}
for $C(q_\perp)$ is used in it.

\begin{figure}
\begin{center}
\includegraphics[width=7cm]{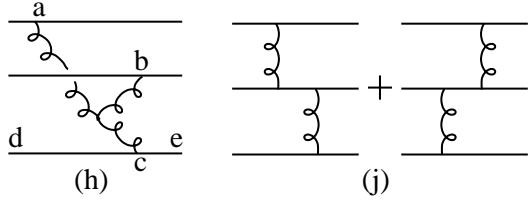}
\end{center}
\caption{Additional diagrams for
the evolution of a triplet of charges.}
\label{fig:threepole}
\end{figure}

\subsection{``Three-pole'' propagation at NLO}

To keep the discussion simple we will assume that particle 3 is a gluon
(color adjoint state), which is sufficient to cover all splitting
processes in QCD (and ${\cal N}=4$ super Yang-Mills).
This ensures that particles 1 and 2 are antiparticles to each other.
We denote by $|s\rangle$ the relevant singlet state in the
tensor product of the three charges;
explicitly, $|s\rangle$ is given by the representation matrices
$(t_1)^a_{ij}$.

The previously treated dipole diagrams (a)-(g) must now be summed
over the three possible
pairs of particles, and we must recompute their group theory factors.
Diagrams (a)-(b) involve, in the case the interaction is 
between particles 1 and 2 \cite{AMYgluon},
\be
-\langle s| t^a_1 \otimes t^a_2 |s\rangle =
\langle s| \frac{t_1^at_1^a+t_2^at_2^a-(t_1{+}t_2)^a(t_1{+}t_2)^a}{2} |s\rangle =
\frac{C_1{+}C_2{-}C_3}{2}, \label{dummygt}
\ee
which reproduces the structure \refeq{gamma3}, upon summing over pairs
and using rotational invariance to ``gauge'' to zero particle 1's
$\perp$-momentum.
Diagrams (c)-(g) fit the same structure, as follows from the
fact that they organize themselves into commutators.
For instance,
\bea
(c) &\propto&
if^{abc} \langle s|t^a_1 t^b_1 \otimes t^c_2 |s\rangle =
-\frac{C_A}{2} \langle s| t^a_1 \otimes t^a_2 |s\rangle,
\nl
(f)+(g) &\propto&
\langle s| [t^a_1,t^b_1] \otimes \frac{[t^a_2,t^b_2]}{2} |s\rangle
=-\frac{C_A}{2} \langle s| t^a_1 \otimes t^{a}_2 |s\rangle.
\eea
Here we have used the identities $[t^a,t^b]=if^{abc}t^c$ and
$f^{abc}f^{abc'}=C_A \delta^{cc'}$.

There are also new diagrams (fig.~\ref{fig:threepole}), which
couple together the three particles nontrivially.
In diagram (h) (see fig.~\ref{fig:threepole}),
the Yang-Mills 3-vertex generates a
factor $f^{abc}$ and the coupling
to the gluon line is given by $(t_3)^{c}_{de}\propto f^{cde}$, whence:
\be
(h)= \langle s| t^a_1 t^b_2 t^c_3 |s\rangle f^{abc} \propto
\Tr_1 \left( t^a t^d t^b t^e\right) f^{abc}f^{dec}= 0\,,
\label{dummygt1}
\ee
with the trace taken in the representation of the particle 1.
We could prove this identity by making extensive use of
the antisymmetry of the $f^{abc}$.
Diagrams (i) are similar to diagram (g) treated in subsection
\ref{sec:diagsdg}, and the main point is that there is a sign
between the two diagrams, due to the reversed middle propagator,
thus yielding zero:
\be
(i)\propto
\langle s| t^a_1 \otimes [t^a_2,t^b_2] \otimes t^b_3 |s\rangle = 0\,.
\ee
Thus the new diagrams (h)-(i) vanish and the factorization formula
\refeq{gamma3} remains valid at NLO.

We view this as somewhat surprising; this could be an
artefact of the
relatively low order in perturbation theory to which we are working.

\subsection{Discussion of operator ordering issues}
\label{sec:oporder}

We now briefly discuss operator ordering issues,
for the Wilson lines in \refeq{dipole}
and their three-particle generalization \refeq{threepole}.
Although this is not directly relevant to the purely classical
effects which are the main object of this paper,
since nonperturbative definitions of $\hat{q}$ have been used
in the literature \cite{teaney} \cite{rajagopal}
we feel that a discussion of them can be of interest.

To help clarify the physical significance of these issues,
let us first consider, in QED, the processes of photon bremsstrahlung
from a charge and of pair production from a photon.
These processes differ in that the former
takes place within the electromagnetic field generated by the 
initial charge, but the latter takes place in an essentially
undisturbed medium (the induced field being suppressed by
the small size of the produced dipole).
The collision kernels relevant to these
two processes could thus be different, due to the different backgrounds,
and should be defined differently.
In the eikonal regime, it is the role of the
Wilson lines trailing behind the charges to source the backgrounds,
which requires that they be properly ordered.

The proper ordering can be readily described using the language of
the Schwinger-Keldysh ``doubled fields'' \cite{schwingerkeldysh},
in which amplitudes and their complex conjugate are described
by type-1 and type-2 fields, respectively.
For photon bremsstrahlung, evolving the relevant $|\psi\gamma\rangle\langle\psi|$
matrix element requires
one type-1 $\psi$ (and $\gamma$) and one type-2 $\psibar$ field,
whereas for pair production,
evolving $|\psi\psibar\rangle\langle \gamma|$
requires both charged fields to be type-1 (and $\gamma$ to be type-2).
In the latter case the Wilson lines nearly cancel against each other
(for a small dipole), whereas in the former case they fail
to cancel, due to operator ordering issues (they live on different branches
of the Keldysh contour): instead they source an electromagnetic field.
This reproduces the expected physics.

The story for QCD must be similar: for instance, evolving 
a $|\psi,g\rangle\langle \psi|$ operator, relevant for gluon
bremsstrahlung,
should require type-1 $\psi$ and $g$ fields, and a type-2 $\psibar$
field, with the obvious replacements to be made for other processes.
Thus we see that the strong coupling calculations of the momentum broadening
coefficient in \cite{teaney} and \cite{gubser}, strictly speaking, gives
a $\hat{q}$ applicable to \emph{photon} bremsstrahlung, whereas the
``jet quenching parameter'' defined in \cite{rajagopal},
being defined from a space-like limit of correlators, is by hypothesis
independent of operator ordering.

It is not clear, at least to the author, the extend to which these
effects can be numerically important.
Obviously, at weak coupling, they are suppressed by a power of the
coupling (the preceding subsection shows that the
suppression is at least $\sim g^2$).
Furthermore, in the $v\nearrow 1$ limit relevant to
high-energy jets, an argument based on the shrinking down of the ``causal diamond''
enclosing any two points on the trajectory of the jet
might suggest that these effects disappear --- e.g. there is no time
available for the induced field to influence the jet back again;
a rigorous analysis, in particular of quantum effects,
will not be attempted here.


\section{Acknowledgements}

I am indebted to Guy~D.~Moore for useful discussions,
and to P.~Arnold for sharing
an early draft of his work \cite{arnoldveryrecent}.
This work was supported in part by
the Natural Sciences and Engineering Research Council of Canada.

\begin{appendix}

\section{Relation to sum rule approach}
\label{app:sumrule}

In this Appendix, we consider the problem of calculating, directly in
four dimensions, the leading order collision kernel \refeq{dipole}:
\be
C(q_\perp)/g^2C_s = \int \frac{dq_z}{2\pi} G_{++}^>(q^0=q_z,q_\perp)
\label{dummya1}
\ee
with $q_\perp\ll T$, and $G_{++}$ the full HTL-resummed
propagator \cite{HTLrefs}.
The simple result \refeq{LOkernel} for it has
been obtained previously using a sum rule by Aurenche,
Gelis and Zaraket (AGZ) \cite{AGZ}, and our aim here is to establish the
equivalence between our approaches.

\subsection{Causality and Sum Rules}
\label{app:sumrule0}

Due to causality, retarded correlators $G^R(Q)$ must be analytic functions
of the four-momentum $Q$ when positive time-like or light-like imaginary
four-vectors are added to it.
This statement extends, in a Lorentz-covariant way,
the familiar analyticity of $G^R$
in the $q^0$ upper-half plane \cite{kapusta}.

Light-like imaginary parts are also allowed, because
causality is preserved along light-fronts (e.g. $G^R(x^+)$ vanishes
for negative light-cone time $x^+$).

In the classical approximation $n_B(q^0)\approx T/q^0$,
\refeq{dummya1} becomes:
\be
\refeq{dummya1}= T \int \frac{dq_z}{2\pi}
\frac{G^R_{++}(q^0=q_z,q_\perp)-G^A_{++}(q^0=q_z,q_\perp)}{q_z}.
\ee
To evaluate this by contour integration, we first move
the $q_z=0$ pole slightly off-axis, 
$1/q_z\to 1/(q_z-i\epsilon)$, which does not change the result
since $(G^R{-}G^A)$ vanishes at $q_z=0$.
Next, we note, using the standard HTL expressions \cite{HTLrefs},
that $G^{R,A}_{++}$ vanishes (like $1/q_z^2$) at large
$|q_z|$, making it possible to close integration contours at infinity.
Closing the contour for $G^R$ (resp. $G^A$)
in the upper (resp. lower) half-plane,
one obtains a unique residue $iT G^R_{++}(q^0=q_z=0,q_\perp)$ from $G^R$
and nothing from $G^A$, due to their aforementioned analyticity
properties, thus reproducing \refeq{LOkernel}:
\be
\refeq{dummya1}=
T\left(\frac{1}{q_\perp^2}-\frac{1}{q_\perp^2{+}m_D^2}\right).
\label{appres}
\ee

Additional poles at the Matsubara frequencies
$q^0=q_z=2\pi i n T$ would have appeared in this result,
in agreement with the sum \refeq{mainres},
had we kept the full Bose  distribution function
$n_B(q^0)$.
This shows that the classical approximation to distribution functions
is equivalent to keeping only the $n=0$ Matsubara frequency.

More generally, at higher orders in perturbation theory
one could also imagine computing the dipole amplitude \refeq{dipole}
using the real-time formalism.
The fact that the operator of interest is supported at a constant
value of $x_+$ allows the integral over the conjugate momentum $q^+$
to be performed by contour integration, in the same way that integrals
over $q^0$
(or equivalently, the sum-integrals of the imaginary-time formalism)
would be done for equal-time correlators \cite{kapusta}.
In addition to standard gauge field propagators, the integrands to be met also
involve eikonal propagators $\sim 1/q^-$; but since they
do not depend on $q^+$ they would not interfere with this calculation.
Thus, as claimed in the text, we can also derive all the results of section
\ref{sec:calc}
using causality-based sum rules.  This is in fact how we first obtained them.
It is also clear that, in general, making the approximation
$n_B(p^0)\to p^0/T$ in all propagators in real-time
is equivalent to dropping all $n\neq 0$ Matsubara modes.

\subsection{AGZ's sum rule}

AGZ \cite{AGZ} study exactly the integral \refeq{dummya1}, but
parametrized using
a different variable, $x=q^0/q$ (so $q^0(x)=q_z(x)=|q_\perp| x/\sqrt{1-x^2}$):
\be
\refeq{dummya1}= |q_\perp| \int_{-1}^1 \frac{dx}{2\pi (1{-}x^2)^{3/2}}
G_{++}^>(x,q_\perp)
\label{dummya2}
\ee
A key observation in \cite{AGZ} is that the HTL propagators, viewed
as a function of $x$ with $q_\perp$ fixed and $q^0=q_z$,
are analytic in the whole complex $x$-plane, apart from a
branch cut at real $x\in[-1,1]$.
Using methods of complex analysis,
they could then derive the result \refeq{appres}.

To show that this analyticity property in $x$ is equivalent to the analyticity
in $q^+$ that we have used above (e.g. to causality),
we rewrite the change of variable above \refeq{dummya2} as:
\be
q^0(x)=q_z(x)= i |q_\perp| \frac{x}{\sqrt{x^2-1}},
\label{dummyx}
\ee
and choose to put the branch cut of the square root at real $x\in
[-1,1]$.
Thus $q^0\to i|q_\perp|$ as $|x|\to\infty$ in any direction.
This choice of branch cut ensures that $G^R(q^0(x),q_z(x),q_\perp)$
goes into the standard
retarded function as ${\rm Im}\,x\to 0^+$,
and is consistent with the conventions of \cite{AGZ}, e.g.
this function has the same analytic structure as the $G^R(x,q_\perp)$
of \cite{AGZ}.
Careful inspection of \refeq{dummyx} then reveals that the imaginary part
of $q^0$ is \emph{positive} for all $x$, establishing that analyticity
in $x$ (for $q^0=q_z$ and at fixed $q_\perp$)
is a consequence of the above-discussed analyticity in $q^+$.
It thus applies to any propagator, extending the claim of \cite{AGZ}.

The authors of \cite{AGZ} worked in the Coulomb gauge
and found, at intermediate steps,
contributions from the large circle at $|x|= \infty$
(proportional to $1/(q_\perp^2{+}\frac13 m_D^2)$),
which in the end, precisely canceled out between the longitudinal and
transverse channels.
Since no such term has appeared in our approach,
the reader might wonder as to their claimed equivalence.
What happens is that these contributions are mere gauge artefacts;
this also explains their ultimate cancellation.
To see this, we note that the residue at $|x|\to\infty$ corresponds to
a pole at $q^0=q_z=iq_\perp$, which is at an ordinary point in the upper-half
$q^+$-plane and is thus forbidden by causality.
But since a gauge like the Coulomb gauge does not respect
causality in a Lorentz-covariant sense
(its $A^0$ field mediates an instantaneous Coulomb interaction),
such poles are not forbidden in individual, gauge-dependent terms.
They are bound, however, to cancel out in physical
quantities like $C(q_\perp)$.
Our approach assumes Lorentz-covariant causality from the start and
cannot detect such unphysical contributions.

These complications, associated with causality violations
in non-covariant gauges (in intermediate expressions),
are easily avoided by working in a covariant gauge.

\end{appendix}

\end{document}